# Can ChatGPT evaluate research environments? Evidence from REF2021

**Kayvan Kousha**
Statistical Cybermetrics and Research Evaluation Group, Business School, University of Wolverhampton, UK. https://orcid.org/0000-0003-4827-971X; Email: k.kousha@wlv.ac.uk

**Mike Thelwall**
School of Information, Journalism and Communication, University of Sheffield, UK. https://orcid.org/0000-0001-6065-205X; Email: m.a.thelwall@sheffield.ac.uk

**Elizabeth Gadd**
Research & Innovation Office, Loughborough University, UK. https://orcid.org/0000-0003-4509-7785; Email: e.a.gadd@lboro.ac.uk

UK academic departments are evaluated partly on the statements that they write about the value of their research environments for the Research Excellence Framework (REF) periodic assessments. These statements mix qualitative narratives and quantitative data, typically requiring time-consuming and difficult expert judgements to assess. This article investigates whether Large Language Models (LLMs) can support the process or validate the results, using the UK REF2021 unit-level environment statements as a test case. Based on prompts mimicking the REF guidelines, ChatGPT 4o-mini scores correlated positively with expert scores in almost all 34 (field-based) Units of Assessment (UoAs). ChatGPT's scores had moderate to strong positive Spearman correlations with REF expert scores in 32 out of 34 UoAs: 14 UoAs above 0.7 and a further 13 between 0.6 and 0.7. Only two UoAs had weak or no significant associations (Classics and Clinical Medicine). From further tests for UoA34, multiple LLMs had significant positive correlations with REF2021 environment scores (all $p < .001$), with ChatGPT 5 performing best ($r=0.81$; $ρ=0.82$), followed by ChatGPT-4o-mini ($r=0.68$; $ρ=0.67$) and Gemini Flash 2.5 ($r=0.67$; $ρ=0.69$). If LLM-generated scores for environment statements are used in future to help reduce workload, support more consistent interpretation, and complement human review then caution must be exercised because of the potential for biases, inaccuracy in some cases, and unwanted systemic effects. Even the strong correlations found here seem unlikely to be judged close enough to expert scores to fully delegate the assessment task to LLMs.

## Introduction

Strong research systems and cultures help researchers work together effectively, access the resources they need, and develop their skills. These factors also support the short- and longer-term production of high-quality research. Analyses of UK Research Excellence Framework (REF) environment statements, which are evidence-based claims of good research support environments, suggest that research environments are believed to be facilitated by strategic coherence, leadership, collaboration, inclusion, and researcher development (Inglis et al., 2024).

Evaluating the research environment is a key part of the UK REF national research assessment. In REF2021, the Environment element contributed 15% of the overall quality scores of each institutional submission. In the forthcoming REF2029, this element was set to expand to "People, Culture and Environment" and carry a higher share of the overall REF score (REF2029, 2025). This change reflects a growing awareness of the importance of good research cultures as an inherent part of any 'excellent' research environment. In REF2021, environment scores represented the proportion of each submission judged to meet quality levels ranging from unclassified to 4*. These scores, together with those for quality of outputs (50%) and impact narratives (25%) produced the overall quality profile for each higher education institution within each broadly field-level Unit of Assessment (UoA). Funding bodies use these outcomes to allocate quality related (QR) research funding.

In the REF context, assessing research environments is complex because excellence can take different forms across disciplines, institutions, and career stages (Inglis, Gadd & Stokoe, 2024). Guidance for REF2021 stated that REF unit-level environment statements would be assessed along the dimensions of "vitality" and "sustainability", using both qualitative narratives and quantitative indicators. This covers organisational culture, impact, people, research infrastructure, income generation, and collaboration (REF2021, 2019a; REF2021, 2019b). The assessment of environment statements is multifaceted and complex, as they combine both qualitative narratives and quantitative indicators (e.g., doctoral degrees awarded, research income) and assess two different dimensions.

Although there has been little prior research about evaluating environment statements, it is known that assessing the somewhat similar impact case studies is time-consuming. For instance, impact assessors spent a median of 85 hours on evaluations, compared with 167 hours for Main Panel Members, and 241 hours for advisers (Manville et al., 2015, p. 59). An independent review of the REF also highlighted "there are some problematic features of the REF which give rise to costs, burdens and distortions which could be reduced or avoided" (Stern, 2016, p. 11). Since unit-level environment statements are usually longer and contain more data (see next section), their evaluation is likely to be more complex and time-consuming. Moreover, evidence from the past UK Research Assessment Exercise (RAE) showed that some universities benefited from a "halo effect" from their reputation and that scoring standards differed between panels, suggesting that expert assessments of research environments may also be uneven and biased (Taylor, 2011). The Metric Tide report also found that the REF2014 environment statements were the "least successful element" in the use of quantitative data. Panels said that narrative environment statements were hard to judge, highlighting difficulties in "separating quality in research environment from quality in writing about it," and they also found some data difficult to interpret without knowing the size or structure of the unit. The report recommended wider use of quantitative indicators in future REF cycles, but only when supported by enough contextual information to ensure consistent interpretation (Wilsdon et al., 2015). These issues suggest that assessing environment statements is challenging and put intensive burden on reviewers, creating a need for methods that support assessment. Hence, using Large Language Models (LLMs) to support the assessment of environment statements might reduce the burden on reviewers, save resources, complement expert judgement, and help achieve greater consistency across evaluations, but only if LLMs can be shown to be capable at this task.

Previous studies using REF2021 data have shown that ChatGPT-based scoring correlate significantly with departmental average scores for both research outputs (Thelwall & Yaghi, 2025b) and impact case studies (Kousha & Thelwall, 2025). Here, the ChatGPT 4o-mini API was prompted with the REF2021 evaluation guidelines and produced scores that correlated positively and weakly to moderately with departmental average expert scores across almost all Units of Assessment. These findings suggest that ChatGPT can provide useful assessments to complement or cross-check academic judgements. The current study investigates whether ChatGPT scores for REF2021 unit-level research environment statements correlate positively with the REF2021 expert score profiles. It also includes further analyses for Unit 34, comparing results from ChatGPT 4o-mini with those generated by ChatGPT 5 and Gemini Flash 2.5 to assess how newer models perform relative to earlier ones.

## REF environment statements

The environment element of REF2021 is designed to assess the *vitality* and *sustainability* of research environments within UK higher education institutions. It focuses on the strategies, people, and resources that enable high-quality and impactful research. In REF2021 institutions were required to submit two environment statements (REF2021, 2019a, REF2021, 2019b): An institutional-level environment statement providing an overview of the institutional research environment and multiple unit-level environment statements which describe the specific research environment for each UoA. The latter statements formed the basis for the assessment of the environment sub-profile.

The unit-level environment statement was required to follow a standardised structure of four sections, as specified in the REF2021 Guidance on Submissions (REF2021, 2019a, p. 100):
- Unit context and structure, research and impact strategy – evidence of the unit's strategic aims, achievements, and future plans for research and impact.
- People – staffing strategy, development and training of postgraduate research students, and equality and diversity support.
- Income, infrastructure and facilities – information on funding, research income, physical and digital resources, and facilities supporting research and impact.
- Collaboration and contribution to the research base, economy and society – evidence of partnerships, networks, and wider contributions to research, policy, and practice.

The unit-level environment statements must include both qualitative narratives and quantitative indicators. These indicators cover aspects such as research income, research student numbers, and research degrees awarded, providing measurable evidence of research environment successes (REF2021, 2019a, p. 101). The four Main Panels requested different information to some extent. For example, Main Panel A (medical and life sciences) emphasized ethics and benchmarked indictors, B (physical sciences and engineering) emphasised strategic investment and quantitative indicators, C (social sciences) emphasised strategy and interdisciplinary collaboration, and D (arts and humanities) emphasised intellectual environments and public engagement.

Panel members assessed the narratives and supporting indicators using the following five-point scale (REF2021, 2019a, p. 85):
- 4* (world-leading): An environment that is conducive to producing research of world-leading quality and enabling outstanding impact, in terms of its vitality and sustainability.
- 3* (internationally excellent): An environment that is conducive to producing research of internationally excellent quality and enabling very considerable impact, in terms of its vitality and sustainability.
- 2* (internationally recognised): An environment that is conducive to producing research of internationally recognised quality and enabling considerable impact, in terms of its vitality and sustainability

- 1* (nationally recognised): An environment that is conducive to producing research of nationally recognised quality and enabling recognised but modest impact, in terms of its vitality and sustainability.
- Unclassified: An environment that is not conducive to producing research of nationally recognised quality or enabling impact of reach and significance.

## Assessing REF environment statements

Several studies have analysed REF environment statements using topic modelling, text analysis, regression and qualitative methods to understand how research environments are described and relate to REF assessment outcomes.

A study of 1,888 REF2021 environment statements used topic modelling and identified 42 topics including eight main themes. These themes associated with about 59% of the variation in panel environment scores, increasing to 69% when research income, doctoral degrees, and staff FTE were added. Thus, panel scores strongly associated with how statements are written (Inglis et al., 2024). Text analysis of 98 REF2014 environment statements in Business and Management found clear tone differences between institutions. Higher-ranked universities used more language indicating stability or confidence, while lower-ranked institutions discussed ongoing development or change (Thorpe et al., 2018a). A follow-up study using the same dataset examined the writing style of environment studies in Business and Management finding clear and consistent style differences between higher and lower scored submissions (Thorpe et al., 2018b).

Equality and diversity have also been studied using environment statements. A HEFCE-commissioned text-mining and content analysis of the 'People' sections from 1,911 REF2014 environment statements showed that most units mainly reported gender-related work, especially the national Athena SWAN initiative. Mentions of other protected characteristics such as ethnicity, disability, age, sexual orientation, and religion or belief were far less common. The study also found "a positive relationship between REF research environment sub-profiles (scores) and reference to key E&D terms within submissions" (Mellors-Bourne et al., 2017, p. 2).

Another quantitative study analysed research income per Full Time Equivalent (FTE) staff member in REF2021 Education (UoA23) using data reported in environment statements. It found moderate positive associations between total income per FTE and overall GPA (r=0.590), with higher correlations for UKRI-related income (r=0.644) and lower correlations for UK government (r=0.295) and EU government income (r=0.244). This suggests that some funding sources may help evidence research excellence in the REF (Boeren, 2023).

In a regression analysis of 1,790 REF2014 environment submissions across 36 Units of Assessment research income per FTE and unit size were significant positive predictors of environment scores in most UoAs (31 and 28 out of 36). Postgraduate research completions had little effect, and Russell Group universities and units with a panel member tended to receive higher scores even after controlling for income and size, suggesting that both resources and institutional reputation influenced evaluations (Pinar & Unlu, 2020).

Some studies have used qualitative methods to investigate how research environments were experienced. A study of research culture and environment based on interviews, workshops, and a survey of 4,267 researchers found concerns about how assessment systems shape working environments. Only 14% felt that current metrics helped to improve research culture, while 43% said their workplace valued metrics more than research quality. Most (54%) reported pressure to meet performance targets such as REF or grant requirements, rising to 63% among academic researchers (Wellcome Trust, 2020, p. 18). Content analysis has also identified disciplinary and institutional differences in how research environments and support activities such as libraries and research services are described in REF environment statements (Pike & Barrantes, 2025; Walker, 2020).

## LLMs for research evaluation

There is substantial evidence that LLMs have some ability to mimic academic evaluation tasks in the sense of giving results that correlate positively with expert judgement. The ICLR conference series has been a pioneer in this area, with reviewers given access to LLM evaluations of submitted conference papers, producing evidence that LLM suggestions can be helpful for reviewers (Liang et al., 2024; Thakkar et al., 2025) and have some ability to predict peer review outcomes (Thelwall & Yaghi, 2025a; Kousha & Thelwall, 2025; Zhou et al., 2024). They can even be used to evaluate reviews and incentivise reviewers (Lu et al., 2024) but also threaten peer review due to reviewers relying on them for their reviewing tasks, rather than just language polishing (Yu et al., 2025).

In terms of research quality scoring and ranking, it is now known that a wide variety of medium sized and larger LLMs have an ability to rank journal articles for quality in a way that correlates positively with expert scores in some or all fields. Most evidence is from ChatGPT 4o-mini, the smaller version of the multimodal ChatGPT 4o, which gives scores that correlate positively with a proxy for REF scores in all fields, although slightly weaker than for ChatGPT 4o (Thelwall & Yang, 2025). The correlations were strongest in the health and physical sciences and weakest in the arts and humanities, and vary little with different LLM settings (Thelwall, 2025b). The correlations are similar if the input is the article full text or just the title and abstract (Thelwall, 2024, 2025b), showing that the scores are pattern-based guesses rather than genuine evaluations. Two key findings from this line of research have been the need to submit a prompt multiple times and use the mean score, and the fact that the score predictions are inaccurate and should not be used, but the rankings based on these predictions are informative.

Other models that also have an ability to score academic journal articles for research quality include Google's cloud-based Gemini (Thelwall, 2025c), and its open weights model Gemma3 (Thelwall, 2025d). In addition, open weights models of different sizes have been tested for related tasks, but not for score-based ranking (Wu et al., 2025).

## Research questions

Since ChatGPT-generated scores correlate significantly with expert departmental averages for both REF2021 research outputs (Thelwall & Yaghi, 2025b) and impact case studies (Kousha & Thelwall, 2025), it is reasonable to believe that large language models might also be able to assess claims about research environments. However, no previous study has tested whether ChatGPT can evaluate unit-level research environment statements. The main difference between these and impact case studies, which are also structured narratives, is that they are supported by standardised quantitative indicators. This study addresses three research questions. The final one is exploratory and relates to only one UoA because it uses labour intensive data gathering.

- **RQ1.** To what extent do ChatGPT scores for REF2021 unit-level environment statements correlate with REF2021 departmental environment average scores?

- **RQ2.** Are there disciplinary differences in the relationship between ChatGPT and REF2021 scores across the 34 Units of Assessment?
- **RQ3.** How do ChatGPT 4o-mini scores compare with those produced by newer models (ChatGPT 5 and Gemini Flash 2.5) in evaluating research environment statements for Unit 34 (Communication, Cultural and Media Studies, Library and Information Management)?

## Methods

This study analysed REF2021 unit-level environment statements available from the public REF database (https://results2021.ref.ac.uk/environment). In total, 1,872 statements were downloaded in PDF format and converted to plain text. All headings, footnotes, and page numbers were removed using a computer program so the data would be consistent before being processed by the ChatGPT 4o-mini API. The environment component of the REF comprises two elements: the narrative and a separate set of quantitative statistics on doctoral degrees awarded (separate by year), research income (15 funder categories, such as "EU government bodies") and research income in kind (2 categories) (available at: https://results2021.ref.ac.uk/environment/export-all). The statistics were not included on the basis that the core information would be referred to and appropriately contextualised in the environment statements, and this is a more natural way for it to be processed by a LLM. Nevertheless, this means that the LLM processes different information than the human REF assessors. UoA assessors were also provided with university-wide institutional environment statements to help contextualise the UoA environment statements. These were not provided to the LLM because they were not scored and were only provided for context, so providing them seems likely to skew the LLM score too much towards institutional-level reputation or quality.

The ChatGPT 4o-mini API was used to generate star ratings for each unit-level environment statement. Following the approach used in earlier studies (Thelwall & Yaghi, 2025b; Kousha & Thelwall, 2025), each statement was fed into the model along with a prompt reflecting the REF2021 environment evaluation criteria. ChatGPT 4o-mini was chosen because it is a relatively cost-effective model with API access for large-scale analysis and has given good results for two other REF components, as reviewed above. It also does not learn from earlier prompts, so each environment statement was scored independently.

The prompt instructed ChatGPT 4o-mini to assign a star rating, using almost the same wording as the REF guidelines (Appendix A). Half-point scores (e.g., 2.5*, 3.5*) were allowed to have a finer-grained assessment, since this approach worked for impact case studies (Kousha & Thelwall, 2025). To reduce random variation in ChatGPT scoring, each statement was evaluated five times independently and the average of these five scores was used as the final score. This approach was based on a previous study (Thelwall & Yaghi, 2025b) which found that averaging at least five iterations of ChatGPT 4o-mini gives stronger associations and more reliable agreement with expert scores than using a single run. ChatGPT 4o-mini was instructed to act as a strict academic expert critically assessing REF environment statements ("*You are a very strict academic expert, critically assessing environment statements […]*") based on REF2021 guidance to reflect the standards and expectations of expert reviewers in the actual REF evaluation process, again following a strategy that worked well for impact case studies (Kousha & Thelwall, 2025). This is because earlier experiments showed that stricter prompting reduced the tendency of ChatGPT-4o-mini to score generously (often awarding 4*) and produced evaluations that more strongly correlated with REF panel average scores. To test whether the two newer models ChatGPT 5 and Gemini Flash 2.5 could give better results, Unit 34 (Communication, Cultural and Media Studies, Library and Information Management) was selected for comparisons. UoA34 was selected because it includes enough environment statements (58) for correlation analysis and is familiar to the authors as their "home" UoA. The unit is also interdisciplinary, combining qualitative and quantitative elements within environment statements. The same strict expert prompt used with ChatGPT 4o-mini was also used for both models through their web interfaces. Each of the 58 statements from this unit was evaluated five times independently in each model and every run was conducted in a new conversation to avoid learning effects from previous conversations. The average of the five scores for each statement was then compared with the environment score for the submission (since the official score is reported as percentages of the environment achieving the levels 4*, 3*, 2*, 1* and 0/U, the percentage-weighted average was used to give a single overall score for comparison) to examine differences between models.

In all cases, the usefulness of the LLM scores was assessed by calculating the Spearman correlation between the REF2021 expert scores and the LLM scores for all statements within a UoA. Pearson correlations are included as an additional descriptive indicator. Spearman correlations are the most important because they assess the rank order agreement, which is the primary value of the scores. Rank order is more important than score predictions because LLM scores for academic documents tend to be inaccurate (e.g., Kousha & Thelwall, 2025; Thelwall & Yaghi, 2025b).

## Results

### Panel-level associations between ChatGPT and expert scores

Across all four REF Main Panels, ChatGPT 4o-mini average scores had strong and statistically significant positive correlations with departmental environment averages (r=0.63 to 0.71, Figure 1). Panel B (Physical Sciences, Engineering & Mathematics) and Panel C (Social Sciences) had the highest correlations (around r = 0.70), so ChatGPT performed particularly well in subjects with an emphasis on quantitative indicators. Panel A (Medicine, Health & Life Sciences) also had a high Pearson correlation (r=0.71), while Panel D (Arts & Humanities) had the lowest but still high correlation (r=0.63). One possible reason is that arts and humanities environment statements rely more on qualitative evidence and narrative explanation rather than quantitative indicators as evidence of achievement (even though the structured quantitative data was not given to the LLM), especially for research funding. Moreover, success at research funding outside the social sciences, arts and humanities seems more likely to translate into spending on the environment, such as for new equipment, support workers and perhaps even buildings to house them. These disciplines often describe culture, collaboration, and impact in more interpretive and subjective ways, perhaps making them less directly comparable to STEM environment statements. This is consistent with previous studies with lower correlations between ChatGPT and expert scores in arts and humanities than in STEM for REF outputs (Thelwall & Yaghi, 2025b) and impact case studies (Kousha & Thelwall, 2025).

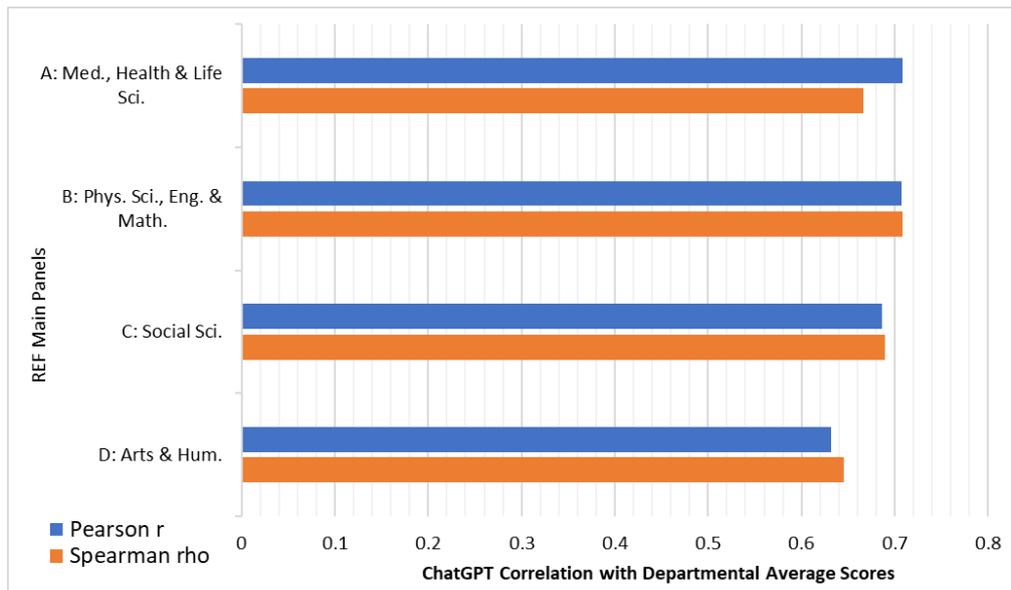

**Figure 1.** Correlations between ChatGPT 4o-mini average scores and REF2021 departmental average environment scores across the four Main Panels (A–D). Both Pearson r and Spearman correlations are statistically significant at p <.001.

## UoA-level correlations between ChatGPT and expert scores

At the Unit of Assessment level, correlations between ChatGPT 4o-mini average scores and REF2021 departmental environment averages were generally strong and statistically significant across most REF subjects, except for Classics and Clinical Medicine (Figure 2). Pearson and Spearman coefficients typically ranged between 0.65–0.80, indicating relatively strong associations between the scores generated by ChatGPT and those assigned by expert assessors.

The highest Spearman correlations occurred in Sociology (ρ = 0.794), Psychology, Psychiatry and Neuroscience (ρ=0.781), Theology and Religious Studies (ρ = 0.761), Archaeology (ρ=0.758), and Engineering (ρ=0.759). Most other UoAs also had moderate to strong positive Spearman associations. By contrast, lower or non-significant correlations appeared in a few disciplines that rely more heavily on qualitative and interpretive evidence. For instance, Classics had no significant relationship, and Clinical Medicine had only a weak Pearson correlation (r=0.419, p=0.019) with no significant Spearman association (ρ=0.301, p=0.10). The weakest but still significant Spearman correlations were in Modern Languages and Linguistics (ρ=0.375, p=0.009), Anthropology and Development Studies (ρ=0.412, p=0.036) and Philosophy (ρ=0.466, p=0.005).

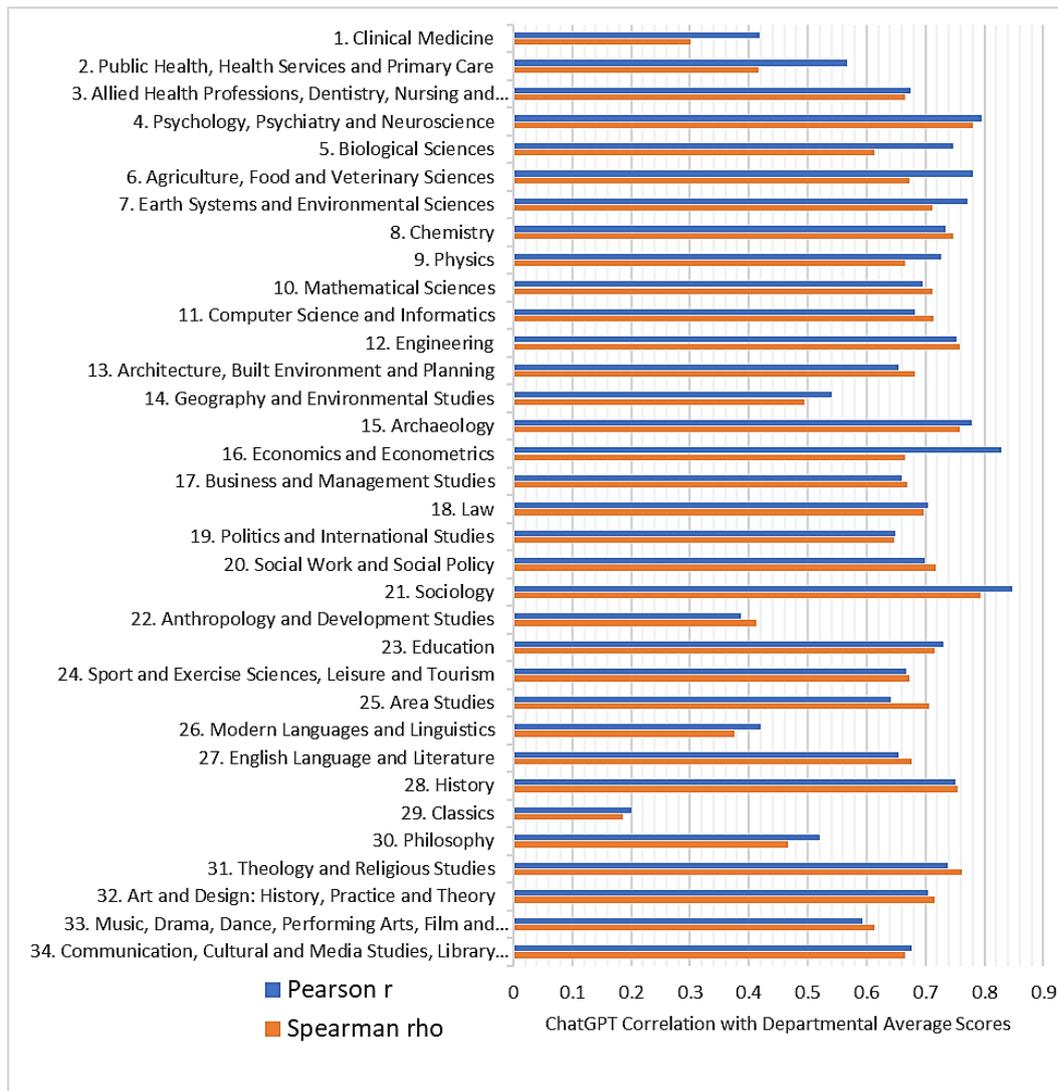

**Figure 2.** Correlations between ChatGPT 4o-mini average scores and REF2021 departmental average environment scores across 34 Units of Assessment. Both Pearson r and Spearman correlations are statistically significant at p < .001 unless stated otherwise.

**Correlations between LLM models and expert scores (Unit 34)**

For Unit 34, all three models (ChatGPT 5, ChatGPT 4o mini, and Gemini Flash 2.5) had statistically significant positive correlations with REF2021 departmental environment scores (Figure 3). The ChatGPT 5 model had the highest correlation with expert scores (Pearson r=0.81; ρ=0.82), followed by ChatGPT 4o mini (r=0.68; ρ=0.67) and Gemini Flash 2.5 (r=0.67; ρ=0.69). All correlations were significant at p < .001, so all models had some ability to capture patterns similar to experts assessing environment statements.

The newest and largest model tested, ChatGPT 5, has the highest level of correlation with expert scores. GPT-5 is claimed by OpenAI to be an improved model for reasoning, factual accuracy, and instruction (OpenAI, 2025).

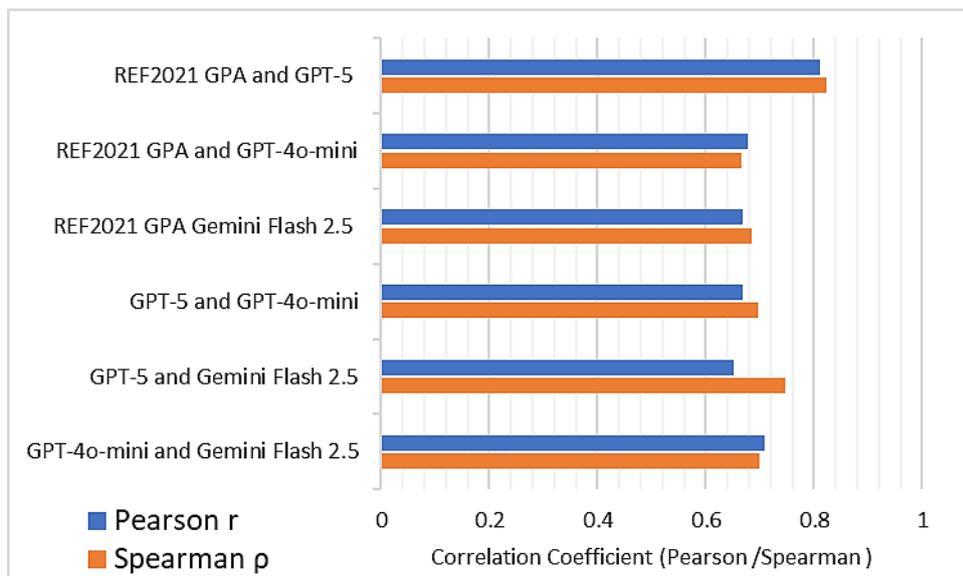

**Figure 3.** Correlations between ChatGPT 4o-mini, ChatGPT 5, Gemini Flash 2.5, and REF2021 departmental environment scores for Unit 34, and inter-model comparisons. All correlations are statistically significant at p < .001.

## Discussion

The results are limited by the use of a single main LLM for the main tests and a single case study, UK REF2021. Although the only large-scale available case to study, REF2021 environment statements are not ideal because they are available online in PDF format, so ChatGPT may have read them, and score profiles for them are available publicly online. Whilst it seems unlikely that ChatGPT would have connected the environment statements submitted to it with the online score profiles, if it had read them, this is a technical possibility. This is because the score profiles are primarily available either in a data-rich visual form (e.g., https://results2021.ref.ac.uk/profiles/units-of-assessment/3) or a data-rich spreadsheet (https://results2021.ref.ac.uk/profiles/export-all), whereas LLMs are primarily trained on unstructured text. They can process spreadsheets but convert them to a text format first, such as in comma separated variable form and are highly capable at processing individual spreadsheets from users but this is different from the task of learning from one spreadsheet amongst a mass of other textual data. An LLM could easily work out that any given text is a REF environment statement and could be reasonably certain of the host university and UoA from the text but connecting it to information extracted from one of the two data-rich sources would be problematic. Moreover, LLMs learn through repeated exposure to related information, so it seems unlikely, but not impossible, that ChatGPT could connect many statements to scores from the two data rich sources, even if it had read them (this is unknown and secret to OpenAI). It seems more possible that it might connect an environment statement to institutional press releases or other institutional promotional online text (e.g., "Clinical Medicine has 100 percent 4* ranking for research environment", https://www.kcl.ac.uk/research/impact/ref) to form positive associations, especially for institutions with a substantial promotional campaign around REF2021 environment scores. Moreover, it may have formed positive associations with the higher-scoring universities or departments through its general training data, and these could have influenced its scoring pattern, despite being explicitly told to score the statement rather than the department.

In terms of disciplinary differences, the low correlation for UoA 29 Classics could be a side effect of its small size (n=17) but this does not explain the anomalously low Spearman correlation for UoA 1 Clinical Medicine (n=31). This UoA has the highest average ChatGPT score but the fourth highest REF2021 score (Figure 4). Overall, 23 out of 31 departments were given the top score 4* by ChatGPT, but only 4 by the REF panel so the underlying cause of the low correlation is ChatGPT giving scores that are too high, not differentiating between excellent departments. It may do this because of the nature of medicine, which involves careful ethics oversight, high external funding for well-resourced studies, and direct impacts on human health. For example, despite being only the third largest UoA by full time equivalent staff submitted, Clinical Medicine had the highest research income overall (£9.9 billion; Engineering was second with £7.1 billion), the second highest income in kind (£1.1 billion, behind Physics with £1.3 billion, presumably due to collaborative high energy physics and astronomy infrastructures), and the second highest number of doctoral degrees (1,738, behind the 52% larger Engineering with 3,656). REF2021 assessors may consciously or subconsciously take UoA-wide high averages into account when norm referencing the UoA, but ChatGPT cannot do this in the way that it was configured for this experiment, with each department being assessed separately.

We also tested whether using five repeated ChatGPT 5.1 scores for each of the 31 departments in Clinical Medicine would improve the association with the experts average REF scores but again found no significant Spearman correlation ($\rho = 0.306$, $p = 0.094$). Hence, even the latest ChatGPT 5.1 model does not reproduce the scoring patterns assigned by REF experts in Clinical Medicine, although ChatGPT 5.1 produced a slightly lower average score (3.8) compared with ChatGPT 4o-mini (3.9). This may be because environment statements in Clinical Medicine often include highly specialised clinical evidence, detailed research structures, and complex institutional narratives that require expert knowledge and interpretation to assess their quality.

A possible reason for relatively low correlations in some UoAs is that they contain both high and low resourced specialties, with departments tending to focus on one or the other. If REF2021 assessors take this into account but ChatGPT doesn't (because it was not told to) then this would disrupt the correlation. This could apply to UoA 26 Modern Languages and Linguistics, which combines the humanities-oriented area of modern languages with the computing-oriented computational linguistics, the latter requiring computing resources and seeming to be better funded.

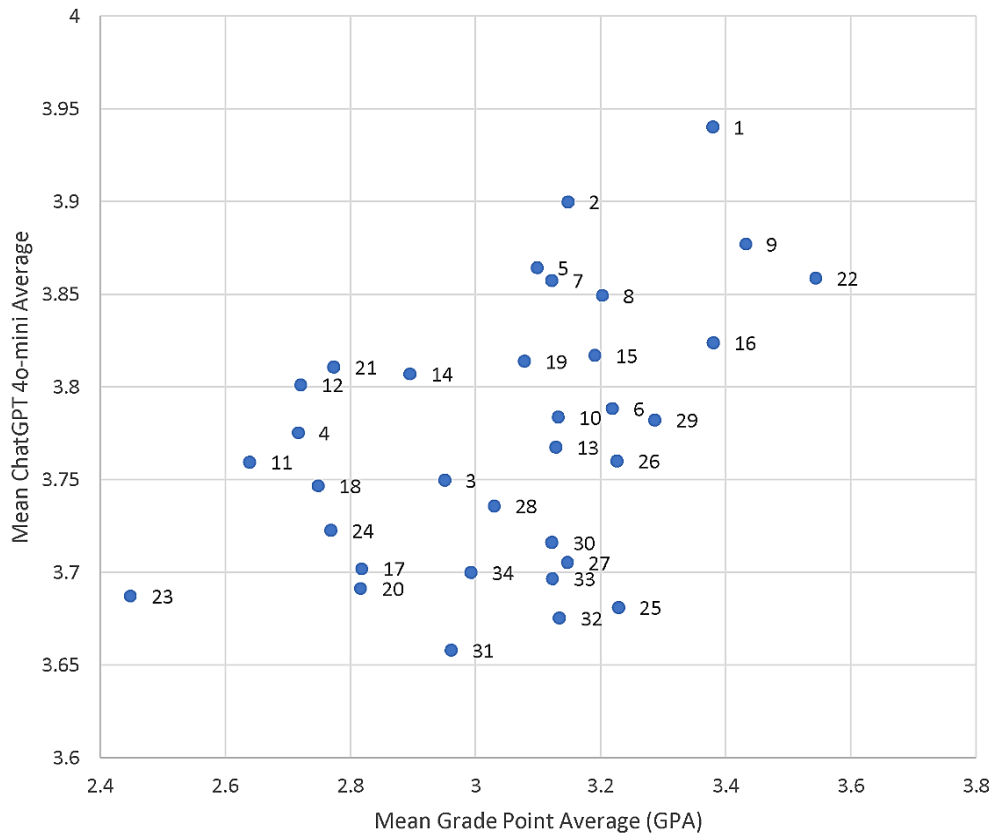

**Figure 4.** Mean ChatGPT 4o-mini score against mean REF2021 score for environment statements in all UoAs. The numbers on the chart refer to the UoA codes (e.g., 1 = Clinical Medicine, 17= Business & Management, as in Figure 2).

All UoAs had a mean score of above 3.65; the modal score was 4*, with 3.5* being low by contrast, which is much higher overall than for the human REF2021 assessors (Figure 4). Thus, ChatGPT was generally much more impressed by the statements than were the human assessors.

The individual ChatGPT reports (see Appendix A and Appendix B for examples) tended to be structured (although in different ways), discussing the research environment and with an explicit, albeit vague, justification for the overall score. Examining the reports, the most evident omission is the lack of norm referencing within a UoA. The human REF2021 assessors would focus on a single UoA and would naturally and probably consciously norm reference at least the quantitative evidence and probably also the narrative claims. Thus, an arguably more natural way to ask AI to assess environment statements would be in complete sets for a UoA, but this would entail an enormous about of text to process for each prompt and does not align with how they work. An alternative strategy to enable a degree of norm referencing might be needed.

## Factors identified by ChatGPT as limiting 4* scores

Across all 29 environment statements in Unit 34 that received an average ChatGPT 5 score of 3.5 or below, the model justified specific weaknesses that prevented a 4* score. An informal content analysis of these showed that funding was mentioned in every case (100%) as a limiting factor, followed by international activity (89.7%), infrastructure and strategy (86.2%), staffing depth (65.5%), impact (69%), and postgraduate research (51.7%) (Table 1).

**Table 1.** Frequency (%) of ChatGPT-identified reasons for not awarding a 4* score in environment statements averaging 3.5 or below (n=29).

| ChatGPT-identified reasons for not awarding 4* | Frequency | % |
| --- | --- | --- |
| Funding/research income | 29 | 100% |
| International activity and engagement | 26 | 89.7% |
| Infrastructure and strategy | 25 | 86.2% |
| Staffing and structure | 19 | 65.5% |
| Impact | 20 | 69% |
| Postgraduate research (PGR) | 15 | 51.7% |

ChatGPT commonly highlighted research income as too low for a world-leading environment. For example, it justified that total income was often "modest", "income per FTE" remained below what it expected for a 4*, or funding relied mainly on small or medium grants, local schemes, a few individuals or not sufficiently competitive at UKRI level. ChatGPT also directly cited specific income amounts for the majority of all cases when explaining why a unit fell short of 4* expectations, such as "total income (£480k) is modest relative to the breadth of activity and staff size" or "the overall research income (£287k over the period) is low in comparison to national benchmarks for a unit of this size and disciplinary spread". This suggests that ChatGPT gives very strong weight to the scale and distribution of research funding when distinguishing between internationally excellent (3*) and world-leading (4*) environments, although it is not clear if it can recognise the different funding norms across disciplines. For instance, the ChatGPT justification that "Research income of £1.1m (including major AHRC, ESRC, EPSRC, and Horizon 2020 awards) demonstrates success in competitive funding, though the scale remains modest for a unit of this size and ambition" is arguable.

ChatGPT often justified limited international activity as a reason for not awarding 4*. It mentioned collaborations that were small-scale or regional rather than global, or dependent on individual staff rather than institutional structures. For staff-related limitations, the model frequently noted reliance on one or two senior academics, uneven contributions across the unit, limited diversity at senior levels, or a staff base that was too small for what it expected of a world-leading environment. Common concerns about postgraduate research included small doctoral cohorts, the absence of externally funded postgraduate researchers, and questions about long-term sustainability. ChatGPT often judged impact activity to fall below 4* level, for example where impacts were mainly regional, relied on individual academics, or were supported by descriptive rather than independently evidenced claims.

## Conclusions

The results show for the first time that AI-based score estimates for environment statements can give plausible results, at least in the sense of tending to correlate at least moderately well with expert scores. Importantly, the individual score predictions are not accurate but the rank order of units based on their environment statements is the useful information produced. The results tend to validate both the expert scores (which are subjective) and ChatGPT (which makes linguistic pattern-based guesses), assuming that they are independent of each other.

Concerns about the use of AI in research evaluation are not new, particularly regarding risks of bias, transparency, and over-reliance on automated judgement (Kousha & Thelwall, 2024). These issues remain relevant when considering how LLMs might support or complement expert assessment of research environments. Therefore, while LLMs may provide useful supplementary insights, it is still premature to replace expert judgements with AI. Instead, LLM-generated scores might be used to cross-check expert assessments for potential biases and to offer informal feedback on draft environment statements. In low resource environments, where evaluation expertise is unavailable, a stronger role might be considered. Of course, if LLMs start to take an important role in evaluating environment statements then systemic effects should be considered (Thelwall, 2025a). A 2025 REF-AI report (Watermeyer et al., 2025) also shows that generative AI is already starting to influence how institutions prepare their REF submissions, with both opportunities and risks. This suggests that while LLMs can provide broadly plausible scores for environment statements with moderate associations with expert assessments, they should be used cautiously, such as for informal feedback during drafting or to help cross-check for possible inconsistencies or biases, rather than to replace expert judgement.

**In answer to the first research question**, ChatGPT scores had moderate to strong positive Spearman correlations with REF2021 expert scores for unit-level environment statements in 32 out of 34 UoAs. Moreover, 27 out of 34 UoAs had correlations above 0.6, suggesting that the ChatGPT might be able to rank the quality of environment statements to a similar degree as academic experts, although the extent to which experts would agree is unknown. This suggests that LLMs could be used to support REF environment assessment by giving a very quick and low-cost evaluation based on the REF guidelines. In this way, LLMs could help with cross-checking or complementing peer review of environment statements. The strength of the correlations also suggests that LLMs might be used to provide early feedback for institutions preparing draft environment statements, helping them identify strengths and weaknesses or complement internal and external review processes in universities before REF expert assessment.

**In answer to the second research question**, the results had clear disciplinary differences. Many UoAs had Spearman correlations above 0.7, including Sociology, Psychology, Theology and Religious Studies, Archaeology, and Engineering. These units might use structures or narratives in their environment statements or include more transparent core evidence, making them easier for LLMs to assess. In contrast, the weakest significant correlations were found in Modern Languages and Linguistics, Anthropology and Development Studies, and Philosophy which may rely more on research culture promotion, which is harder to evaluate. The weak correlation for Clinical Medicine may be because all the environment statements are impressive in this area and LLMs are reluctant to give any a score below the maximum.

**In answer to the third research question,** all three models (ChatGPT 4o-mini, ChatGPT 5, and Gemini Flash 2.5) tested on UoA 34 environment statement had significant and similar positive correlations with the expert environment scores. This suggests that the ability to partially mimic the human assessment of environment statements is widespread and core to LLM technology, rather than an artefact of any particular one.

# References

Boeren, E. (2023). Investigating the relationship between research income and research excellence in education: Evidence from the REF2021-UoA23 data. British Educational Research Journal, 49(6), 1312-1337.

Inglis, M., Gadd, E., & Stokoe, E. (2024). What is a high-quality research environment? Evidence from the UK's research excellence framework. Research Evaluation, rvae010. https://doi.org/10.1093/reseval/rvae010

Kousha, K., & Thelwall, M. (2024). Artificial intelligence to support publishing and peer review: A summary and review. Learned Publishing, 37(1), 4-12. https://doi.org/10.1002/leap.1570

Kousha, K., & Thelwall, M. (2025). Assessing the societal influence of academic research with ChatGPT: Impact case study evaluations. Journal of the Association for Information Science and Technology, 76(10), 1357-1373. https://doi.org/10.1002/asi.25021

Liang, W., Zhang, Y., Cao, H., Wang, B., Ding, D. Y., Yang, X., & Zou, J. (2024). Can large language models provide useful feedback on research papers? A large-scale empirical analysis. *NEJM AI*. https://doi.org/10.1056/AIoa2400196

Lu, Y., Xu, S., Zhang, Y., Kong, Y., & Schoenebeck, G. (2024). Eliciting informative text evaluations with large language models. In *Proceedings of the 25th ACM conference on economics and computation* (pp. 582–612).

Manville, C., Guthrie, S., Henham, ML., Garrod, B., Sousa, S., Kirtley, A., et al. (2015). Assessing impact submissions for REF 2014: An evaluation. Rand Europe. https://www.rand.org/content/dam/rand/pubs/research_reports/RR1000/RR1032/RAND_RR1032.pdf

Mellors-Bourne, R., Metcalfe, J., & Gill, A. (2017). Exploring Equality and Diversity using REF2014 Environment Statements. CRAC LTD Report to HEFCE. https://vitae.ac.uk/wp-content/uploads/2025/02/Exploring-equality-and-diversity-using-REF2014-environment-statements.pdf

OpenAI. (2025, August 7). Introducing GPT 5. Retrieved from https://openai.com/index/introducing-gpt-5/

Pike, C., & Barrantes, B. S. L. (2025). Libraries as research enablers: A content analysis of research excellence framework environment statements. The Journal of Academic Librarianship, 51(3), 103039.

Pinar, M., & Unlu, E. (2020). Determinants of quality of research environment: An assessment of the environment submissions in the UK's Research Excellence Framework in 2014. Research Evaluation, 29(3), 231-244.

REF2021 (2019a). Guidance on submissions (2019/01). https://2021.ref.ac.uk/publications-and-reports/guidance-on-submissions-201901/index.html

REF2021 (2019b). Panel criteria and working methods (2019/02). https://2021.ref.ac.uk/publications-and-reports/panel-criteria-and-working-methods-201902/index.html

REF2029 (2025). REF2029 Guidance. https://2029.ref.ac.uk/guidance/

Stern, N. (2016). Building on Success and Learning from Experience: An Independent Review of the Research Excellence Framework. Department for Business, Energy & Industrial Strategy. https://assets.publishing.service.gov.uk/media/5a803df4e5274a2e8ab4f03d/ind-16-9-ref-stern-review.pdf

Taylor, J. (2011). The assessment of research quality in UK universities: Peer review or metrics?. British Journal of Management, 22(2), 202-217.

Thakkar, N., Yuksekgonul, M., Silberg, J., Garg, A., Peng, N., Sha, F., & Zou, J. (2025). Can LLM feedback enhance review quality? a randomized study of 20k reviews at ICLR 2025. arXiv preprint arXiv:2504.09737.

Thelwall, M. & Yaghi, A. (2025a). Evaluating the predictive capacity of ChatGPT for academic peer review outcomes across multiple platforms. Scientometrics. https://doi.org/10.1007/s11192-025-05287-1

Thelwall, M. & Yaghi, A. (2025b). In which fields can ChatGPT detect journal article quality? An evaluation of REF2021 results. Trends in Information Management, 13(1), 1-29.

Thelwall, M. & Yang, Y. (2025). Implicit and explicit research quality score probabilities from ChatGPT. *Quantitative Science Studies*. https://doi.org/10.1162/QSS.a.393

Thelwall, M. (2024). Can ChatGPT evaluate research quality? Journal of Data and Information Science, 9(2), 1–21. https://doi.org/10.2478/jdis-2024-0013

Thelwall, M. (2025a). Research quality evaluation by AI in the era of Large Language Models: Advantages, disadvantages, and systemic effects. Scientometrics. https://doi.org/10.1007/s11192-025-05361-8

Thelwall, M. (2025b). Evaluating research quality with Large Language Models: An analysis of ChatGPT's effectiveness with different settings and inputs. Journal of Data and Information Science, 10(1), 7-25. https://doi.org/10.2478/jdis-2025-0011

Thelwall, M. (2025c). Is Google Gemini better than ChatGPT at evaluating research quality? Journal of Data and Information Science, 10(2), 1–5. https://doi.org/10.2478/jdis-2025-0014 with extended version here: https://doi.org/10.6084/m9.figshare.28089206.v1

Thelwall, M. (2025d). Can smaller large language models evaluate research quality? Malaysian Journal of Library and Information Science, 30(2), 66-81. https://doi.org/10.22452/mjlis.vol30no2.4

Thorpe, A., Craig, R., Hadikin, G., & Batistic, S. (2018a). Semantic tone of research 'environment' submissions in the UK's Research Evaluation Framework 2014. Research Evaluation, 27(2), 53-62.

Thorpe, A., Craig, R., Tourish, D., Hadikin, G., & Batistic, S. (2018b). 'Environment' Submissions in the UK's Research Excellence Framework 2014. British Journal of Management, 29(3), 571-587.

Walker, D. (2020). Libraries and the REF: how do librarians contribute to research excellence? Insights, 33(6).

Watermeyer, R., Phipps, L., Benites, R., & Crick, T. (2025). REF-AI: Exploring the potential of generative AI for REF2029. Research England. https://bpb-eu-w2.wpmucdn.com/blogs.bristol.ac.uk/dist/3/1073/files/2025/11/Full-report-final.pdf

Wellcome Trust. (2020). What researchers think about the culture they work in. https://cms.wellcome.org/sites/default/files/what-researchers-think-about-the-culture-they-work-in.pdf

Wilsdon, J., Allen, L., Belfiore, E., Campbell, P., Curry, S., Hill, S., Jones, R., Kain, R., Kerridge, S., Thelwall, M., Tinkler, J., Viney, I., Wouters, P., & Hill, J. (2015). The Metric Tide: Report of the Independent Review of the Role of Metrics in Research Assessment and Management. HEFCE. https://www.ukri.org/wp-content/uploads/2021/12/RE-151221-TheMetricTideFullReport2015.pdf

Wu, W., Zhang, Y., Haunschild, R., & Bornmann, L. (2025). Leveraging Large Language Models for post-publication peer review: Potential and limitations. In: Editors: Shushanik Sargsyan, Wolfgang Glänzel, Giovanni Abramo: 20th International Conference on Scientometrics & Informetrics, 23–27 June 2025, Yerevan, Armenia, (p. 1176-1195).

Yu, S., Luo, M., Madusu, A., Lal, V., & Howard, P. (2025). Is Your Paper Being Reviewed by an LLM? Benchmarking AI Text Detection in Peer Review. *arXiv preprint arXiv:2502.19614*.

Zhou, R., Chen, L., & Yu, K. (2024). Is LLM a reliable reviewer? A comprehensive evaluation of LLM on automatic paper reviewing tasks. In Proceedings of the 2024 joint international conference on computational linguistics, language resources and evaluation (LREC-COLING 2024) (pp. 9340–9351).

# Appendix A

*System ChatGPT prompt used for scoring REF unit-level environment statements*

You are a very strict academic expert, critically assessing environment statements describing the vitality and sustainability of research environments. You will provide a score of 1* to 4* alongside a detailed justification.

For the purposes of this assessment, the research environment is defined as "its vitality and sustainability, including the approach to enabling impact from its research, and its contribution to the vitality and sustainability of the wider discipline or research base."

Environment statements should include information about "context and mission, strategy, people, and income, infrastructure and facilities." They should also show "how the unit supports collaboration both within and beyond academia, and where these arrangements have enhanced the vitality and sustainability of the research environment or the unit's contribution to the wider research base, economy or society."

The scoring system used is 1*, 2*, 3* or 4*, which are defined as follows:

4*: An environment that is conducive to producing research of world-leading quality and enabling outstanding impact, in terms of its vitality and sustainability.

3*: An environment that is conducive to producing research of internationally excellent quality and enabling very considerable impact, in terms of its vitality and sustainability.

2*: An environment that is conducive to producing research of internationally recognised quality and enabling considerable impact, in terms of its vitality and sustainability.

1*: An environment that is conducive to producing research of nationally recognised quality and enabling recognised but modest impact, in terms of its vitality and sustainability.

Use half points if an environment statement is between two scores

You will understand vitality as the degree to which the research environment is active, dynamic and conducive to supporting high-quality research, including the approach to enabling impact from its research.

You will understand sustainability as the extent to which the research environment has the people, strategy, income, infrastructure, facilities and collaborations needed to maintain and develop its vitality into the future, and to contribute to the vitality and sustainability of the wider discipline, research base, economy or society.

You will make an overall judgement about the vitality and sustainability of the research environment, rather than assessing each element separately. While environment statements should demonstrate evidence across areas such as context, mission, strategy, people, income, infrastructure and facilities, and collaboration, the balance between them may vary at all quality levels. You will exercise your judgement without privileging or disadvantaging any single aspect.

## Appendix B: A 4.0 ChatGPT report

Based on the detailed assessment of the environment statement from Cardiff University's Unit of Assessment 1 (UOA1) School of Medicine (SoM), the following score and justification are provided:

**Score: 4.0**

**Justification:**

1. **Context and Structure:**

   - The UOA1 has established a strong and coherent context within the broader School of Medicine and the College of Biomedical and Life Sciences. The focus on multidisciplinary collaboration, especially in precision medicine and experimental research, underlines its vitality.

2. **Research and Impact Strategy:**

   - The strategy devised since the last REF (2014) clearly articulates goals that support interdisciplinary research, significantly enhancing research performance and productivity. Achievements such as over 2,900 articles published and securing substantial grants (£3148.8 million) are indicative of a high-impact research environment.

3. **People:**

   - The UOA1 places a clear emphasis on equality, diversity, and the development of personnel. The initiatives aimed at enhancing the career opportunities of early-career researchers (ECRs), coupled with significant recruitment in strategic areas and promotion structures, show a commitment to nurturing talent and sustaining a vibrant research community.

4. **Income, Infrastructure, and Facilities:**

   - The unit has demonstrated a significant increase in research income (68% rise from REF2014 to REF2021) with a healthy portfolio of diverse funding sources, including government, charity, and industry. The investment in advanced infrastructures (e.g., supercomputing capabilities, specialized imaging facilities) showcases a robust commitment to support high-quality research.

5. **Collaboration and Engagement:**

   - The environment statement highlights extensive collaborations with NHS partners, industry stakeholders, and international research institutions. The establishment of the Clinical Innovation Partnership and the MediCentre business incubator exemplifies the unit's focus on translating research into clinical innovation and societal benefit, thereby enhancing its impact beyond academia.

6. **Governance and Policies:**

   - The governance mechanisms and policies around research integrity, impact monitoring, and ethical standards demonstrate a well-structured environment that is conducive to maintaining high standards of research quality.

7. **Sustainability:**

   - The proposal for an integrated research and impact strategy for the future is comprehensive, focusing on interdisciplinary approaches and systemic improvements that address chronic health challenges. Plans to enhance collaboration, resource allocation, and EDI initiatives signal a forward-thinking framework for continued vitality.

The overall impression is of a dynamic and well-resourced research environment that not only produces high-quality research outputs but also fosters a culture of collaboration, inclusivity, and societal impact. Given these strengths, this environment is classified as "conducive to producing research of world-leading quality and enabling outstanding impact".

# Appendix C. A 3.5 ChatGPT report

### Assessment of the Research Environment Statement

#### Overall Score: **3.5*** (Internationally Excellent with Some Aspects of World-Leading Quality)

#### Justification:

**Vitality:**

The [redacted] demonstrates significant vitality through its dynamic research environment. The establishment of the [redacted] as a new entity since 2014 showcases a transformative expansion in STEM fields. The clear research strategy targeting interdisciplinary collaboration among distinct teams, including the [redacted], highlights a proactive approach to fostering innovative research niches. The ambitious research themes spanning healthcare, sustainable materials, and advanced computational methodologies are further supported through operational and resource-rich contexts, including the purpose-built [redacted] building.

Moreover, there is a strong record of publicly recognized achievements, such as the [redacted] project's accolades and significant research awards attributed to individual faculty members. The unit's engagement in high-impact research, such as the development of [redacted] for healthcare, reinforces its active stance in addressing global challenges.

**Sustainability:**

The sustainability of the research environment is evidenced by a robust staffing strategy, ensuring the recruitment of both experienced and early-career researchers, which enriches the research portfolio with diverse skills and perspectives. The substantial investment in research facilities, including access to high-performance computing resources and state-of-the-art wet-lab space, enhances the capacity for ongoing and future research endeavors. The proactive approach to diversifying income streams, with reported external funding exceeding £[redacted] million, indicates a strategic commitment to ensuring financial stability and growth.

Collaboration with high-profile industrial and academic partners, as epitomized in collaborations associated with [redacted], strengthens the unit's position within the wider research ecosystem. The structured outreach efforts that engage PGR students with real-world applications and industry placements ensure a continuous loop of knowledge transfer and relevance to societal needs, bolstering the sustainability of the research narrative.

**Contribution to the Wider Research Base:**

The contributions to the wider research base are well-articulated, showcasing collaborations that extend across national and international boundaries. The significant involvement in community outreach activities, public lectures, and engagement with learned societies further exemplifies [redacted]'s commitment to disseminating findings and fostering an impactful presence in the broader academic community.

The research environment effectively encourages interdisciplinary collaboration, with several initiatives directly linking research in physics to healthcare and sustainable materials, reiterating how [redacted] is not only a participant but a contributor to innovative solutions that address societal challenges.

**Conclusion:**

Despite the commendable aspects outlined, the unit does not fully achieve "world-leading" status due to the relatively nascent development phase it is still undergoing, with ongoing establishment efforts still being part of its narrative. While definite use of impactful research with significant industry and community ties is evident, the potential for greater achievements may require longer-term stability and additional accolades to fully elevate the unit's credibility to the top tier.

The environment is rigorously structured and provides a solid foundation for future growth, with clear objectives and strategic directions laid out for the next five years. [redacted] is positioned to produce even higher-quality research that aligns with its national and global contexts, sustaining its promising trajectory toward becoming a leader in computational physics and interdisciplinary applications.